\documentclass[english,pra,aps,twocolumn,superscriptaddress]{revtex4-2}
\usepackage[T1]{fontenc}
\usepackage[latin9]{inputenc}
\usepackage{xcolor}
\usepackage{mathrsfs}
\usepackage{amsmath}
\usepackage{amssymb}
\usepackage{graphicx}
\PassOptionsToPackage{normalem}{ulem}
\usepackage{ulem}

\makeatletter

\providecolor{lyxadded}{rgb}{0,0,1}
\providecolor{lyxdeleted}{rgb}{1,0,0}
\DeclareRobustCommand{\mklyxadded}[1]{\textcolor{lyxadded}\bgroup#1\egroup}
\DeclareRobustCommand{\mklyxdeleted}[1]{\textcolor{lyxdeleted}\bgroup\mklyxsout{#1}\egroup}
\DeclareRobustCommand{\mklyxsout}[1]{\ifx\\#1\else\sout{#1}\fi}

\usepackage{color}
\usepackage[colorlinks,citecolor=blue,linkcolor=blue]{hyperref}

\makeatother

\usepackage{babel}
\begin{document}
\title{Frequency upconversion of infrared signals via molecular optomechanical
cavities}
\author{Fen Zou}
\affiliation{Center for Theoretical Physics \& School of Physics and Optoelectronic
Engineering, Hainan University, Haikou 570228, China}
\author{Shu-Xian Quan}
\affiliation{Center for Theoretical Physics \& School of Physics and Optoelectronic
Engineering, Hainan University, Haikou 570228, China}
\author{Yong Li}
\email{yongli@hainanu.edu.cn}

\affiliation{Center for Theoretical Physics \& School of Physics and Optoelectronic
Engineering, Hainan University, Haikou 570228, China}
\author{Hui Dong}
\email{hdong@gscaep.ac.cn}

\affiliation{Graduate School of China Academy of Engineering Physics, Beijing 100193,
China}
\date{\today}
\begin{abstract}
Molecular optomechanical cavities have recently emerged as a promising
platform for frequency upconversion, enabling the quantum coherent
conversion of infrared signal into the visible range. In a recent
work {[}F. Zou \textit{et al}., Phys. Rev. Lett. \textbf{132}, 153602
(2024){]}, we proposed an amplification mechanism that can enhance
the intensity of the upconverted infrared signals by a factor of 1000
or more within such a cavity under the ideal case without any noise.
In this work, we employ the power spectrum method to investigate the
noise added to the upconverted signal in a molecular optomechanical
cavity along with the conversion efficiency from infrared signal into
visible range. In the red-detuned regime, the anti-Stokes sideband
achieves superior conversion efficiency relative to the Stokes sideband.
Conversely, the Stokes sideband dominates under the blue-detuned condition,
which amplifies the infrared signal. We further demonstrate the dependence
of the added noise on the coupling strength and decay rates of the
system. In particular, we find that when the infrared signal is amplified,
the added noise approaches the quantum limit of one quantum.
\end{abstract}
\maketitle

\section{Introduction}

The mid-infrared (MIR) or far-infrared (FIR) frequency ranges, spanning
several to hundreds of terahertz (wavelengths from 2.5 to 500~$\mu$m),
constitute a crucial part of the electromagnetic spectrum and underpin
a broad range of applications. These include molecular analysis of
biological tissues~\citep{DeBruyne2018}, clinical medicine~\citep{DeBruyne2018},
thermal imaging~\citep{Tonouchi2007}, quantum sensing~\citep{Kutas2020},
microscopy~\citep{Kviatkovsky2020,Paterova2020}, astronomy~\citep{Ariyoshi2016,Roellig2020},
and homeland security. Infrared (IR) detectors are crucial to these
applications. Conventional IR detectors are highly sensitive to thermal
noise and typically require cryogenic cooling to mitigate thermal
interference. This need for cryogenic cooling makes these detectors
both costly and less sensitive compared with visible (VIS) {[}or near-infrared
(NIR){]} detectors. These limitations highlight the urgent need for
IR detection technologies that are compact, highly sensitive, and
low-noise, while operating without complex cryogenic systems.

A promising approach for detecting MIR and FIR signals is to upconvert
them into the VIS (or NIR) range, where compact, cost-effective, and
highly sensitive cameras are readily available. Existing coherent
conversion methods include nonlinear interferometry~\citep{kalashnikov2016infrared,Lindner2021Nonlinear}
and frequency upconversion~\citep{barh2019parametric,Roelli2020Molecular,Chen2021Continuous,Xomalis2021Detecting,zou2024Amplifying}.
In nonlinear interferometry, IR signals are upconverted into VIS (or
NIR) light through three-wave mixing in bulk nonlinear crystals. And
its efficient upconversion requires stringent phase matching among
the IR, pump, and upconverted fields during propagation~\citep{Temporao2006,Tseng2018}.
Inspired by advances in cavity optomechanics that enable coherent
frequency conversion~\citep{Tian2010,Dong2012,Hill2012,Palomaki2013,Wang2012,Tian2012,Bochmann2013,Andrews2014,Nunnenkamp2014,Metelmann2014,Metelmann2015,Ruesink2018,Shen2018,Forsch2020},
another approach for IR signal detection employs plasmonic nanocavities
containing molecules with both infrared-absorption and Raman-active
vibrational modes to upconvert MIR (or FIR) signals into the VIS (or
NIR) range~\citep{Roelli2020Molecular}. Subsequently, experiments
demonstrated that plasmonic nanocavities containing hundreds of biphenyl-4-thiol
molecules can achieve IR-to-VIS conversion at room temperature~\citep{Chen2021Continuous,Xomalis2021Detecting}.
Yet the low conversion efficiency poses a significant challenge for
the applications of such mechanism of upconversion.

To enhance conversion efficiency, we recently proposed an amplification
mechanism based on molecular collective mode and Stokes sideband pumping
in the molecular optomechanical system~\citep{zou2024Amplifying}.
This scheme has the potential to increase the intensity by a factor
of 1000 or more for the IR signal without considering the impact of
noise. A more comprehensive investigation is needed to evaluate the
impact of system noise on IR signal detection and to quantify the
resulting output noise levels.

In this work, we employ the power spectrum method~\citep{Clerk2010,Malz2018}
to investigate the properties of the upconverted IR signals in a molecular
optomechanical cavity under both red- and blue-detuned conditions,
where the driving frequency equals the sum or difference of the visible
and vibrational frequencies. This method allows us to analyze the
conversion efficiency from IR signal to VIS range, as well as the
noise added to the upconverted signal. We analyze the conversion efficiencies
at both the first Stokes and anti-Stokes sidebands of the pump field.
These results indicate that under the red-detuned condition, the anti-Stokes
sideband exhibits higher conversion efficiency than the Stokes sideband;
the opposite occurs under the blue-detuned condition, where amplification
of the IR signal is achieved. Additionally, we analyze how the system
parameters affect both the conversion efficiency and the noise added
to the upconverted signal. It is shown that the added noise is suppressed
by appropriately tuning the coupling strength and decay rates of the
system. These findings provide a theoretical foundation for future
experimental advances in IR signal upconversion.

\section{Molecular optomechanical system}

\begin{figure}
\centering
\includegraphics{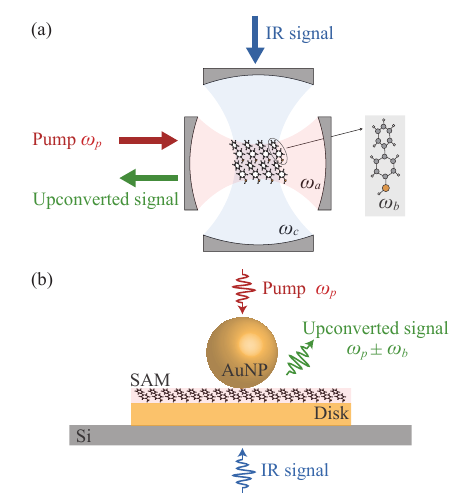}\caption{(a) Schematic diagram of a molecular optomechanical system, where
the molecular vibrations are coupled to the VIS (IR) mode of the cavity
via dispersive (electric-dipole) interaction. A strong pump field
with frequency $\omega_{p}$ is applied to drive the VIS mode within
the cavity. (b) Schematic diagram of a nanoparticle-on-resonator configuration.
From top to bottom, it consists of an Au nanoparticle (AuNP), a self-assembled
monolayer (SAM) of Biphenyl-4-thiol, an Au disk, and a Si layer.}
\label{Fig1}
\end{figure}

We consider a molecular optomechanical system~\citep{Roelli2016Molecular,Jakob2023,Boehmke2024,Huang2024,Roelli2024,Schmidt2024,Abutalebi2024,Kalarde2025,Berinyuy2025,Berinyuy2025a,Peng2025,Huang2025,Munir2025,Yin2025,Yu2026,Berinyuy2026,Tang2025,Berinyuy2025b}
consisting of $N$ identical molecules and a plasmonic nanocavity
that supports both the VIS and IR modes {[}Fig.~\ref{Fig1}(a){]}.
As depicted in Fig.~\ref{Fig1}(b), the plasmonic nanocavity is realized
through a nanoparticle-on-resonator configuration, where an Au nanoparticle
is positioned on top of an Au disk, with the $N$ identical molecules
placed in the gap between the nanoparticle and the disk~\citep{Xomalis2021Detecting}.
In the low-excitation limit of the molecular vibrations, the vibrational
mode of each molecule is approximately modeled as a harmonic oscillator\,\citep{Roelli2016Molecular}.
To achieve frequency upconversion of the IR signal of interest, particular
types of molecules, e.g., Biphenyl-4-thiol, ~\citep{Chen2021Continuous,Xomalis2021Detecting}
are selected to couple simultaneously to both the VIS and IR modes
within the plasmonic nanocavity. A strong pump field in the visible
range, with frequency $\omega_{p}$ and amplitude $\varepsilon_{p}$,
is applied to drive the VIS mode within the nanocavity. The Hamiltonian
of the system is given as ($\hbar=1$)~\citep{zou2024Amplifying}
\begin{align}
H_{s} & =\omega_{a}a^{\dagger}a+\omega_{c}c^{\dagger}c+\sum_{j=1}^{N}\omega_{b}b_{j}^{\dagger}b_{j}+\sum_{j=1}^{N}g_{a}a^{\dagger}a(b_{j}^{\dagger}+b_{j})\nonumber \\
 & \quad+\sum_{j=1}^{N}g_{c}(c^{\dagger}+c)(b_{j}^{\dagger}+b_{j})+i(\varepsilon_{p}e^{-i\omega_{p}t}a^{\dagger}-\text{H.c.}),\label{eq:Hamsys}
\end{align}
where $a$ ($a^{\dagger}$) and $c$ ($c^{\dagger}$) represent the
annihilation (creation) operators for the VIS and IR modes with resonance
frequencies $\omega_{a}$ and $\omega_{c}$, respectively. The operator
$b_{j}$ ($b_{j}^{\dagger}$) is the annihilation (creation) operator
for the vibrational mode of the $j$-th molecule with resonance frequency
$\omega_{b}$. The parameter $g_{a}$ denotes the optomechanical coupling
strength between the VIS mode of the cavity and the molecular vibrations,
arising from the dispersive interaction mediated by the Raman polarizability~\citep{Roelli2016Molecular,Benz2016,Schmidt2016,Lombardi2018,Zhang2020,Esteban2022,Xu2022}.
In contrast, $g_{c}$ represents the bilinear coupling strength between
the IR mode and the molecular vibrations, which originates from the
electric-dipole interaction~\citep{Shalabney2015,Roelli2020Molecular,PannirSivajothi2022}.
For simplicity but without loss of generality, we assume that the
parameters $g_{a}$, $g_{c}$, and $\varepsilon_{p}$ are real. Instead
of adding explicit form of IR signal of interest~\citep{zou2024Amplifying},
we will treat the IR signal as part of the input of the cavity IR
mode following the standard power spectrum methods~\citep{Clerk2010,Malz2018}
in the following discussions.

We introduce the molecular collective operator $B=\sum_{j=1}^{N}b_{j}/\sqrt{N}$,
which satisfies the bosonic commutation relation $[B,B^{\dagger}]=1$.
In the interaction picture with respect to $\omega_{p}a^{\dagger}a$,
Hamiltonian~(\ref{Fig1}) is simplified as
\begin{align}
H & =\Delta_{0}a^{\dagger}a+\omega_{c}c^{\dagger}c+\omega_{b}B^{\dagger}B+G_{a}a^{\dagger}a(B^{\dagger}+B)\nonumber \\
 & \quad+G_{c}(c^{\dagger}+c)(B^{\dagger}+B)+i(\varepsilon_{p}a^{\dagger}-\text{H.c.)},\label{eq:Ham}
\end{align}
where $\Delta_{0}=\omega_{a}-\omega_{p}$ is the detuning between
the VIS mode and the pump field. The parameters $G_{a}=g_{a}\sqrt{N}$
and $G_{c}=g_{c}\sqrt{N}$ represent the collective optomechanical
coupling strength and the collective bilinear coupling strength, respectively.

By substituting Hamiltonian~(\ref{eq:Ham}) into the Heisenberg equation
and accounting for the damping and the corresponding noise terms,
the quantum Langevin equations (QLEs) are obtained as follows:
\begin{align}
\dot{a} & =-(i\Delta_{0}+\kappa_{a})a-iG_{a}a(B^{\dagger}+B)+\varepsilon_{p}+\sqrt{2\kappa_{a}}a_{\text{in}},\nonumber \\
\dot{c} & =-(i\omega_{c}+\kappa_{c})c-iG_{c}(B^{\dagger}+B)+\sqrt{2\kappa_{c}}c_{\text{in}},\\
\dot{B} & =-(i\omega_{b}+\gamma_{B})B-iG_{a}a^{\dagger}a-iG_{c}(c^{\dagger}+c)+\sqrt{2\gamma_{B}}B_{\text{in}},\nonumber 
\end{align}
where $\kappa_{a}$ and $\kappa_{c}$ are the decay rates of the VIS
and IR modes, respectively, and $\gamma_{B}$ is the decay rate of
the molecular collective mode. Note that we consider a single-sided
cavity damping scenario, and neglect the intrinsic decay of both the
VIS and IR modes. The operators $a_{\text{in}}$, $c_{\text{in}}$,
and $B_{\text{in}}=\sum_{j=1}^{N}b_{j,\text{in}}/\sqrt{N}$ represent
the inputs, including both vacuum thermal noise and possible external
coherent input. In the absence of coherent input, these operators
have zero mean values $\langle o_{\text{in}}\rangle=0$ and nonzero
correlation functions given by $\langle o_{\text{in}}^{\dagger}(t)o_{\text{in}}(t^{\prime})\rangle=n_{o}^{\text{th}}\delta(t-t^{\prime})\approx0$
and $\langle o_{\text{in}}(t)o_{\text{in}}^{\dagger}(t^{\prime})\rangle=(n_{o}^{\text{th}}+1)\delta(t-t^{\prime})\approx\delta(t-t^{\prime})$
for $o=a,c,B$. Here we assume the thermal occupations $n_{a,c,B}^{\text{th}}=1/[\exp(\hbar\omega_{a,c,b}/k_{\text{B}}T)-1]$
of the VIS and IR modes as well as the molecular vibrational mode
are negligible due to $\hbar\omega_{a,c,b}/k_{\text{B}}T\gg1$, where
$k_{\text{B}}$ is the Boltzmann constant and $T$ is the ambient
temperature (e.g. $300$ K), i.e., $n_{a,c,B}^{\text{th}}\approx0$.

Here, we would like to remark on the treatment of the input $c_{\mathrm{in}}$.
Without any coherent input, we treat $c_{\mathrm{in}}$ as the vacuum
thermal noise with zero mean values $\langle c_{\text{in}}\rangle=0$
to obtain the steady state. And the later response in the visible
signal is then evaluated by treating the input $c_{\mathrm{in}}$
as the sum of the coherent input and vacuum thermal noise.

Under the condition of a strong pump field, the steady-state mean
values of the operators are obtained using the mean-field approximation,
where $\left\langle a^{\dagger}a\right\rangle _{\mathrm{ss}}\approx\left\langle a^{\dagger}\right\rangle _{\mathrm{ss}}\left\langle a\right\rangle _{\mathrm{ss}}$,
yielding
\begin{align}
\langle a\rangle_{\text{ss}} & =\frac{\varepsilon_{p}}{i\Delta+\kappa_{a}},\nonumber \\
\langle c\rangle_{\text{ss}} & =-\frac{iG_{c}(\langle B\rangle_{\text{ss}}^{*}+\langle B\rangle_{\text{ss}})}{i\omega_{c}+\kappa_{c}},\nonumber \\
\langle B\rangle_{\text{ss}} & =-\frac{i[G_{a}|\langle a\rangle_{\text{ss}}|^{2}+G_{c}(\langle c\rangle_{\text{ss}}^{*}+\langle c\rangle_{\text{ss}})]}{i\omega_{b}+\gamma_{B}},
\end{align}
where $\Delta=\Delta_{0}+G_{a}(\langle B\rangle_{\text{ss}}^{*}+\langle B\rangle_{\text{ss}})$
is the effective detuning.

With the strong pump field, each operator is rewritten as the sum
of the steady-state mean value and the quantum fluctuation, i.e.,
$o=\langle o\rangle_{\text{ss}}+\delta o$ for $o=a,c,B$~\citep{Weis2010}.
By keeping only the first-order term of the quantum fluctuation, the
linearized QLEs are obtained as
\begin{align}
\delta\dot{a} & =-(i\Delta+\kappa_{a})\delta a-i\mathcal{G}_{a}(\delta B^{\dagger}+\delta B)+\sqrt{2\kappa_{a}}a_{\text{in}},\nonumber \\
\delta\dot{c} & =-(i\omega_{c}+\kappa_{c})\delta c-iG_{c}(\delta B^{\dagger}+\delta B)+\sqrt{2\kappa_{c}}c_{\text{in}},\nonumber \\
\delta\dot{B} & =-(i\omega_{b}+\gamma_{B})\delta B-i(\mathcal{G}_{a}^{*}\delta a+\mathcal{G}_{a}\delta a^{\dagger})\nonumber \\
 & \quad-iG_{c}(\delta c+\delta c^{\dagger})+\sqrt{2\gamma_{B}}B_{\text{in}},\label{eq:LinQLEs}
\end{align}
where $\mathcal{G}_{a}=G_{a}\langle a\rangle_{\text{ss}}$ represents
the enhanced collective optomechanical coupling strength. According
to Eq.~(\ref{eq:LinQLEs}), the linearized effective Hamiltonian
is written as
\begin{align}
H_{\text{eff}} & =\Delta\delta a^{\dagger}\delta a+\omega_{c}\delta c^{\dagger}\delta c+\omega_{b}\delta B^{\dagger}\delta B+\mathcal{G}_{a}\delta a^{\dagger}(\delta B^{\dagger}+\delta B)\nonumber \\
 & \quad+\mathcal{G}_{a}^{*}\delta a(\delta B^{\dagger}+\delta B)+G_{c}(\delta c+\delta c^{\dagger})(\delta B^{\dagger}+\delta B).\label{eq:EffHam}
\end{align}
In the following, we use the power spectrum method~\citep{Clerk2010,Malz2018}
to solve the linearized QLEs~(\ref{eq:LinQLEs}).

By defining the fluctuation vector $V=(\delta a,\delta c,\delta B,\delta a^{\dagger},\delta c^{\dagger},\delta B^{\dagger})^{\text{T}}$
and the input field vector $V_{\text{in}}=(a_{\text{in}},c_{\text{in}},B_{\text{in}},a_{\text{in}}^{\dagger},c_{\text{in}}^{\dagger},B_{\text{in}}^{\dagger})^{\text{T}}$,
the linearized QLEs~(\ref{eq:LinQLEs}) is written as
\begin{equation}
\frac{dV}{dt}=-MV+LV_{\text{in}},\label{eq:LinQLEsMat}
\end{equation}
where $L=\text{diag}(\sqrt{2\kappa_{a}},\sqrt{2\kappa_{c}},\sqrt{2\gamma_{B}},\sqrt{2\kappa_{a}},\sqrt{2\kappa_{c}},\sqrt{2\gamma_{B}})$
is the damping matrix and $M=\left(\begin{array}{cc}
P & Q\\
Q^{*} & P^{*}
\end{array}\right)$ is the coefficient matrix with
\begin{equation}
P=\left(\begin{array}{ccc}
i\Delta+\kappa_{a} & 0 & i\mathcal{G}_{a}\\
0 & i\omega_{c}+\kappa_{c} & iG_{c}\\
i\mathcal{G}_{a}^{*} & iG_{c} & i\omega_{b}+\gamma_{B}
\end{array}\right)\label{eq:PMatrix}
\end{equation}
and
\begin{equation}
Q=\left(\begin{array}{ccc}
0 & 0 & i\mathcal{G}_{a}\\
0 & 0 & iG_{c}\\
i\mathcal{G}_{a} & iG_{c} & 0
\end{array}\right).
\end{equation}
The system is stable only when the real parts of all the eigenvalues
of the coefficient matrix $M$ are positive, as determined by the
Routh-Hurwitz criterion~\citep{DeJesus1987,Gradshteyn2014}.

\begin{figure}
\centering
\includegraphics[scale=0.8]{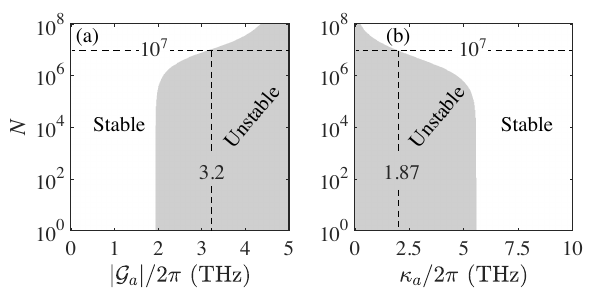}\caption{The stability diagram with respect to (a) the enhanced collective
optomechanical coupling strength $\left|\mathcal{G}_{a}\right|$ and
the number of the molecules $N$ at $\kappa_{a}/2\pi=30\,\text{THz}$;
(b) the decay rate of the VIS mode $\kappa_{a}$ and the number of
the molecules $N$ at $|\mathcal{G}_{a}|/2\pi=0.75\,\text{THz}$.
Here we take the blue-detuned case of $\Delta=-\omega_{b}$. Other
parameters are $\omega_{b}=\omega_{c}=2\pi\times30\,\text{THz}$,
$g_{c}/2\pi=0.1\,\text{GHz}$, $\varepsilon_{p}/2\pi=500\,\text{THz}$,
$\gamma_{B}/2\pi=0.1\,\text{THz}$, and $\kappa_{c}/2\pi=0.5\,\text{THz}$.}
\label{Fig2}
\end{figure}

Figure~\ref{Fig2} illustrates the stability diagram of the system
for the blue-detuned pump field with $\Delta=-\omega_{b}$ and the
resonance condition with $\omega_{c}=\omega_{b}$. Other parameters
are as shown in the caption of Fig.~\ref{Fig2}. The results show
that the system is stable in the white area. For the red-detuned pump
field with $\Delta=\omega_{b}$, the system is entirely stable across
the same parameter range (not shown here).

With the Fourier transform of the operators
\begin{align}
\tilde{o}(\omega) & =\frac{1}{\sqrt{2\pi}}\int_{-\infty}^{+\infty}o(t)e^{-i\omega t}dt,\nonumber \\
\widetilde{o^{\dagger}}(\omega) & =\frac{1}{\sqrt{2\pi}}\int_{-\infty}^{+\infty}o^{\dagger}(t)e^{-i\omega t}dt,
\end{align}
satisfying $[\tilde{o}(\omega)]^{\dagger}=\widetilde{o^{\dagger}}(-\omega)$,
the solution of Eq.~(\ref{eq:LinQLEsMat}) in the frequency domain
is 
\begin{equation}
\tilde{V}(\omega)=(M+i\omega I_{6})^{-1}L\tilde{V}_{\text{in}}(\omega),\label{eq:Solut}
\end{equation}
where $I_{6}$ represents the $6\times6$ identity matrix, and $\tilde{V}(\omega)$
{[}$\tilde{V}_{\text{in}}(\omega)${]} denotes the Fourier transform
of the vector $V$ ($V_{\text{in}}$). By substituting the input-output
relation $\tilde{V}_{\text{out}}(\omega)+\tilde{V}_{\text{in}}(\omega)=L\tilde{V}(\omega)$~\citep{Gardiner1985}
into Eq.~(\ref{eq:Solut}), the output field vector is obtained as
\begin{equation}
\tilde{V}_{\text{out}}(\omega)=U(\omega)\tilde{V}_{\text{in}}(\omega),\label{eq:OutputVec}
\end{equation}
where the scattering matrix $U(\omega)$ is given by 
\begin{equation}
U(\omega)=L(M+i\omega I_{6})^{-1}L-I_{6}.\label{eq:ScaMat}
\end{equation}
The matrix element $U_{jj^{\prime}}(\omega)$ is the element in the
$j$-th row and $j^{\prime}$-th column of the scattering matrix $U(\omega)$,
and denotes the scattering amplitude from the $j^{\prime}$-th component
to the $j$-th one.

In the presence of coherent input, the correlation functions of the
operators $o_{\mathrm{in}}$ ($o=a,c,B$) in the frequency domain
are given by~\citep{Agarwal2012,Xu2015} 
\begin{subequations}
\label{eq:=000020CorrFun}
\begin{align}
\langle\widetilde{o^{\dagger}}_{\text{in}}(\omega)\tilde{o}_{\text{in}}(\omega^{\prime})\rangle & =S_{o,\text{in}}(\omega)\delta(\omega+\omega^{\prime}),\\
\langle\tilde{o}_{\text{in}}(\omega)\widetilde{o^{\dagger}}_{\text{in}}(\omega^{\prime})\rangle & =\left[S_{o,\text{in}}(\omega)+1\right]\delta(\omega+\omega^{\prime}),
\end{align}
\end{subequations}
where $S_{o,\text{in}}(\omega)$ ($o=a,c,B$) denotes the spectrum
of coherent input, and the term ``1'' accounts for the contribution
from the vacuum noise. The output spectrum $S_{o,\text{out}}(\omega)$
is defined by~\citep{Clerk2010}

\begin{equation}
S_{o,\text{out}}(\omega)=\frac{1}{2}\int_{-\infty}^{\infty}\langle\widetilde{o^{\dagger}}_{\text{out}}(\omega)\tilde{o}_{\text{out}}(\omega^{\prime})+\tilde{o}_{\text{out}}(\omega)\widetilde{o^{\dagger}}_{\text{out}}(\omega^{\prime})\rangle d\omega^{\prime}.\label{eq:OutputSpec}
\end{equation}
By substituting Eqs.~(\ref{eq:OutputVec}) and (\ref{eq:=000020CorrFun})
into Eq.~(\ref{eq:OutputSpec}), the output spectrum is given by
\begin{equation}
S_{\text{out}}(\omega)=T(\omega)\left[S_{\text{in}}(\omega)+\frac{1}{2}\left(\begin{array}{c}
1\\
1\\
1
\end{array}\right)\right],
\end{equation}
where $S_{\text{out}}(\omega)=(S_{a,\text{out}}(\omega),S_{c,\text{out}}(\omega),S_{B,\text{out}}(\omega))^{\text{T}}$
and $S_{\text{in}}(\omega)=(S_{a,\text{in}}(\omega),S_{c,\text{in}}(\omega),S_{B,\text{in}}(\omega))^{\text{T}}$.
The matrix $T(\omega)$ is defined as
\begin{equation}
T(\omega)=\left(\begin{array}{ccc}
T_{aa}(\omega) & T_{ac}(\omega) & T_{aB}(\omega)\\
T_{ca}(\omega) & T_{cc}(\omega) & T_{cB}(\omega)\\
T_{Ba}(\omega) & T_{Bc}(\omega) & T_{BB}(\omega)
\end{array}\right),
\end{equation}
where the matrix element $T_{oo^{\prime}}(\omega)$ $\left(o,o^{\prime}=a,c,B\right)$
represents the scattering probability from mode $o^{\prime}$ to mode
$o$ with 
\begin{subequations}
\begin{align}
T_{aa}(\omega) & =\left|U_{11}(\omega)\right|^{2}+\left|U_{14}(\omega)\right|^{2},\\
T_{ac}(\omega) & =\left|U_{12}(\omega)\right|^{2}+\left|U_{15}(\omega)\right|^{2},\label{eq:ConEffic}\\
T_{aB}(\omega) & =\left|U_{13}(\omega)\right|^{2}+\left|U_{16}(\omega)\right|^{2},\\
T_{ca}(\omega) & =\left|U_{21}(\omega)\right|^{2}+\left|U_{24}(\omega)\right|^{2},\\
T_{cc}(\omega) & =\left|U_{22}(\omega)\right|^{2}+\left|U_{25}(\omega)\right|^{2},\\
T_{cB}(\omega) & =\left|U_{23}(\omega)\right|^{2}+\left|U_{26}(\omega)\right|^{2},\\
T_{Ba}(\omega) & =\left|U_{31}(\omega)\right|^{2}+\left|U_{34}(\omega)\right|^{2},\\
T_{Bc}(\omega) & =\left|U_{32}(\omega)\right|^{2}+\left|U_{35}(\omega)\right|^{2},\\
T_{BB}(\omega) & =\left|U_{33}(\omega)\right|^{2}+\left|U_{36}(\omega)\right|^{2}.
\end{align}
\end{subequations}

Considering the coherent input (i.e., the IR signal) to the cavity
IR mode while other inputs without external coherent input, i.e.,
$S_{a,\text{in}}(\omega)=S_{B,\text{in}}(\omega)=0$, the output spectrum
of the VIS mode is given by
\begin{equation}
S_{a,\text{out}}(\omega)=T_{ac}(\omega)\left[S_{c,\text{in}}(\omega)+\frac{1}{2}\right]+S_{a,\text{add}}(\omega),
\end{equation}
where the conversion efficiency $T_{ac}(\omega)$ from the IR signal
to the VIS range is defined in Eq.~(\ref{eq:ConEffic}). The noise
spectrum $S_{a,\text{add}}(\omega)$ added to the upconverted signal,
arising from the vacuum noises of the VIS mode and the molecular vibrational
mode, is given by $S_{a,\text{add}}(\omega)=\left[T_{aa}(\omega)+T_{aB}(\omega)\right]/2$.
Here we show explicitly the matrix elements $U_{1j}(\omega)$ ($j=1,2,...,6$)
as follows
\begin{align}
U_{11}(\omega) & =\{[(\kappa_{c}+i\omega)^{2}+\omega_{c}^{2}]\{[(\Delta+i\kappa_{a})^{2}-\omega^{2}][(\omega-i\gamma_{B})^{2}\nonumber \\
 & \quad-\omega_{b}^{2}]+4\left|\mathcal{G}_{a}\right|^{2}\omega_{b}(\Delta+i\kappa_{a})\}+4G_{c}^{2}\omega_{b}\omega_{c}\nonumber \\
 & \quad\times[(\Delta+i\kappa_{a})^{2}-\omega^{2}]\}F^{-1}(\omega),\nonumber \\
U_{12}(\omega) & =\frac{4G_{c}\mathcal{G}_{a}\sqrt{\kappa_{a}\kappa_{c}}\omega_{b}(\Delta-\omega+i\kappa_{a})[\kappa_{c}+i(\omega-\omega_{c})]}{F(\omega)},\nonumber \\
U_{13}(\omega) & =2i\mathcal{G}_{a}\sqrt{\kappa_{a}\gamma_{B}}(\Delta-\omega+i\kappa_{a})(\omega-\omega_{b}-i\gamma_{B})\nonumber \\
 & \quad\times[(\omega-i\kappa_{c})^{2}-\omega_{c}^{2}]F^{-1}(\omega),\nonumber \\
U_{14}(\omega) & =4i\mathcal{G}_{a}^{2}\kappa_{a}\omega_{b}[(\kappa_{c}+i\omega)^{2}+\omega_{c}^{2}]F^{-1}(\omega),\nonumber \\
U_{15}(\omega) & =\frac{4G_{c}\mathcal{G}_{a}\sqrt{\kappa_{a}\kappa_{c}}\omega_{b}(\Delta-\omega+i\kappa_{a})[\kappa_{c}+i(\omega+\omega_{c})]}{F(\omega)},\nonumber \\
U_{16}(\omega) & =2i\mathcal{G}_{a}\sqrt{\kappa_{a}\gamma_{B}}(\Delta-\omega+i\kappa_{a})(\omega+\omega_{b}-i\gamma_{B})\nonumber \\
 & \quad\times[(\omega-i\kappa_{c})^{2}-\omega_{c}^{2}]F^{-1}(\omega),
\end{align}
where $F(\omega)=\{(\gamma_{B}^{2}+2i\gamma_{B}\omega)[\Delta^{2}+(\kappa_{a}+i\omega)^{2}]-4\left|\mathcal{G}_{a}\right|^{2}\Delta\omega_{b}\}[(\kappa_{c}+i\omega)^{2}+\omega_{c}^{2}]+[\Delta^{2}+(\kappa_{a}+i\omega)^{2}]\{(\omega_{b}^{2}-\omega^{2})[\omega_{c}^{2}+(\kappa_{c}+i\omega)^{2}]-4G_{c}^{2}\omega_{b}\omega_{c}\}$.

At the first anti-Stokes sideband of the pump field, corresponding
to the frequency $\omega_{p}+\omega_{b}$, the IR-to-VIS conversion
efficiency is given by $T_{ac}^{AS}=T_{ac}(\omega=-\omega_{b})$~\citep{note},
and the noise added to the upconverted signal is expressed as~\citep{Nunnenkamp2014,Malz2018,jiang2018}
\begin{equation}
n_{\text{add}}^{AS}(\omega)=\frac{S_{a,\text{add}}(\omega)}{T_{ac}^{AS}}.
\end{equation}
Similarly, at the first Stokes sideband of the pump field, with frequency
$\omega_{p}-\omega_{b}$~\citep{note}, the IR-to-VIS conversion
efficiency is $T_{ac}^{S}=T_{ac}(\omega=\omega_{b})$, and the added
noise is given by
\begin{equation}
n_{\text{add}}^{S}(\omega)=\frac{S_{a,\text{add}}(\omega)}{T_{ac}^{S}}.
\end{equation}
\begin{figure}
\centering
\includegraphics{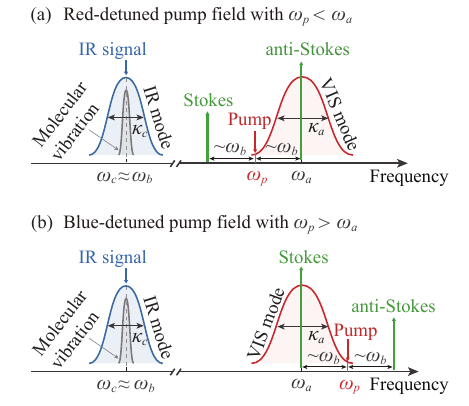}\caption{(a, b) Schematic diagram illustrating the upconversion of the IR signal
to the VIS range within a molecular optomechanical system under the
red-detuned and blue-detuned pump fields. It should be noted that
the IR signal is treated as part of the input $c_{\text{in}}$ of
cavity IR mode.}
\label{Fig3}
\end{figure}

\section{Red-detuned pump field\protect\label{sec:RedDetunedPumpField}}

In this section, we will analytically derive the conversion efficiency
from the IR signal to the VIS range under red-detuned pump field {[}see
Fig.~\ref{Fig3}(a){]} with the rotating-wave approximation (RWA),
i.e., $\left\{ \left|\mathcal{G}_{a}\right|,G_{c}\right\} \ll\Delta\sim\omega_{b}\sim\omega_{c}$.

Under the above RWA conditions, the fast-oscillating terms $(\mathcal{G}_{a}\delta a^{\dagger}\delta B^{\dagger}+\text{H.c.})$
and $(G_{c}\delta B^{\dagger}\delta c^{\dagger}+\text{H.c.})$ in
Eq.~(\ref{eq:EffHam}) can be neglected. The linearized QLEs in Eq.~(\ref{eq:LinQLEs})
is obtained as,
\begin{align}
\delta\dot{a} & =-(i\Delta+\kappa_{a})\delta a-i\mathcal{G}_{a}\delta B+\sqrt{2\kappa_{a}}a_{\text{in}},\nonumber \\
\delta\dot{c} & =-(i\omega_{c}+\kappa_{c})\delta c-iG_{c}\delta B+\sqrt{2\kappa_{c}}c_{\text{in}},\label{eq:RedLinQLEs}\\
\delta\dot{B} & =-(i\omega_{b}+\gamma_{B})\delta B-i\mathcal{G}_{a}^{*}\delta a-iG_{c}\delta c+\sqrt{2\gamma_{B}}B_{\text{in}}.\nonumber 
\end{align}
By defining the fluctuation vector $\mathcal{V}=(\delta a,\delta c,\delta B)^{T}$
and the input field vector $\mathcal{V}_{\text{in}}=(a_{\text{in}},c_{\text{in}},B_{\text{in}})^{T}$,
the linearized QLEs~(\ref{eq:RedLinQLEs}) is written as
\begin{equation}
\frac{d\mathcal{V}}{dt}=-P\mathcal{V}+\mathcal{J}\mathcal{V}_{\text{in}},
\end{equation}
where the coefficient matrix $P$ is given in Eq.~(\ref{eq:PMatrix}),
and the damping matrix $\mathcal{J}$ is given by $\mathcal{\mathcal{J}}=\text{diag}(\sqrt{2\kappa_{a}},\sqrt{2\kappa_{c}},\sqrt{2\gamma_{B}})$.

In the case of only the IR signal in the input, the output spectrum
of the VIS mode in the frequency domain is given by
\begin{equation}
S_{a,\text{out}}(\omega)=\mathcal{T}_{ac}^{\mathrm{rd}}(\omega)\left[S_{c,\text{in}}(\omega)+\frac{1}{2}\right]+S_{a,\text{add}}(\omega),
\end{equation}
where the conversion efficiency $\mathcal{T}_{ac}^{\mathrm{rd}}(\omega)$
from the IR signal to the VIS range is expressed as
\begin{align}
\mathcal{T}_{ac}^{\mathrm{rd}}(\omega) & =\left|\mathcal{U}_{12}^{\mathrm{rd}}(\omega)\right|^{2}.\label{eq:RedConEffic}
\end{align}
The added noise spectrum is given by $S_{a,\text{add}}(\omega)=\left[\mathcal{T}_{aa}^{\mathrm{rd}}(\omega)+\mathcal{T}_{aB}^{\mathrm{rd}}(\omega)\right]/2$,
where $\mathcal{T}_{aa}^{\mathrm{rd}}(\omega)=\left|\mathcal{U}_{11}^{\mathrm{rd}}(\omega)\right|^{2}$
and $\mathcal{T}_{aB}^{\mathrm{rd}}(\omega)=\left|\mathcal{U}_{13}^{\mathrm{rd}}(\omega)\right|^{2}.$
Here, $\mathcal{U}_{1j}^{\mathrm{rd}}(\omega)$ denotes the element
in the first row and $j$-th column of the scattering matrix $\mathcal{U}^{\mathrm{rd}}(\omega)=\mathcal{\mathcal{J}}(P+i\omega I_{3})^{-1}\mathcal{\mathcal{J}}-I_{3}$,
with $\mathcal{U}_{12}^{\mathrm{rd}}(\omega)=-2G_{c}\mathcal{G}_{a}\sqrt{\kappa_{a}\kappa_{c}}/\{G_{c}^{2}(i\Delta+i\omega+\kappa_{a})+[\left|\mathcal{G}_{a}\right|^{2}+(i\Delta+i\omega+\kappa_{a})(i\omega_{b}+i\omega+\gamma_{B})](i\omega_{c}+i\omega+\kappa_{c})\}$
.

As shown in Fig.~\ref{Fig3}(a), we consider the red-detuned pump
field with $\Delta=\omega_{b}$ and the resonance case with $\omega_{c}=\omega_{b}$.
At the first anti-Stokes sideband of the pump field, i.e., $\omega_{p}+\omega_{b}$,
the conversion efficiency from the IR signal to the VIS range is given
by
\begin{equation}
\mathcal{T}_{ac}^{\mathrm{rd},AS}\equiv\mathcal{T}_{ac}^{\mathrm{rd}}(\omega=-\omega_{b})=\left|\frac{-2G_{c}\mathcal{G}_{a}\sqrt{\kappa_{a}\kappa_{c}}}{G_{c}^{2}\kappa_{a}+\left|\mathcal{G}_{a}\right|^{2}\kappa_{c}+\kappa_{a}\kappa_{c}\gamma_{B}}\right|^{2}.\label{eq:RedASConEff}
\end{equation}
When $\kappa_{a}\kappa_{c}\gamma_{B}\ll\{\left|\mathcal{G}_{a}\right|^{2}\kappa_{c},G_{c}^{2}\kappa_{a}\}$,
the conversion efficiency $\mathcal{T}_{ac}^{\mathrm{rd},AS}$ approaches
1 under the condition $\left|\mathcal{G}_{a}\right|^{2}\kappa_{c}\simeq G_{c}^{2}\kappa_{a}$.

At the first Stokes sideband of the pump field, i.e., $\omega_{p}-\omega_{b}$,
the conversion efficiency from the IR signal to the VIS range is
\begin{equation}
\mathcal{T}_{ac}^{\mathrm{rd},S}\equiv\mathcal{T}_{ac}^{\mathrm{rd}}(\omega=\omega_{b})=\left|\frac{-2G_{c}\mathcal{G}_{a}\sqrt{\kappa_{a}\kappa_{c}}}{G_{c}^{2}\Gamma_{a}+\left|\mathcal{G}_{a}\right|^{2}\Gamma_{c}+\Gamma_{a}\Gamma_{c}\Gamma_{B}}\right|^{2},\label{eq:RedSConEff}
\end{equation}
where $\Gamma_{a}=2i\omega_{b}+\kappa_{a}$, $\Gamma_{c}=2i\omega_{b}+\kappa_{c}$,
and $\Gamma_{B}=2i\omega_{b}+\gamma_{B}$.

\section{Blue-detuned pump field}

In Sec.~\ref{sec:RedDetunedPumpField}, we analyzed the conversion
efficiency from the IR signal to the VIS range under the red-detuned
pump field condition. In this section, we will extend the discussion
to the conversion efficiency for a blue-detuned pump field {[}see
Fig.~\ref{Fig3}(b){]}. Under the RWA conditions $\left\{ \left|\mathcal{G}_{a}\right|,G_{c}\right\} \ll-\Delta\sim\omega_{b}\sim\omega_{c}$,
the fast-oscillating terms $(\mathcal{G}_{a}\delta a^{\dagger}\delta B+\text{H.c.})$
and $(G_{c}\delta B^{\dagger}\delta c^{\dagger}+\text{H.c.})$ in
Eq.~(\ref{eq:EffHam}) is usually considered to be negligible. Hence,
we obtain the linearized QLEs as
\begin{align}
\delta\dot{a} & =-(i\Delta+\kappa_{a})\delta a-i\mathcal{G}_{a}\delta B^{\dagger}+\sqrt{2\kappa_{a}}a_{\text{in}},\nonumber \\
\delta\dot{c}^{\dagger} & =-(-i\omega_{c}+\kappa_{c})\delta c^{\dagger}+iG_{c}\delta B^{\dagger}+\sqrt{2\kappa_{c}}c_{\text{in}}^{\dagger},\nonumber \\
\delta\dot{B}^{\dagger} & =-(-i\omega_{b}+\gamma_{B})\delta B^{\dagger}+i\mathcal{G}_{a}^{*}\delta a+iG_{c}\delta c^{\dagger}+\sqrt{2\gamma_{B}}B_{\text{in}}^{\dagger}.\label{eq:BlueLinQLEs}
\end{align}
By defining the fluctuation vector $\mathscr{\mathcal{W}}=(\delta a,\delta c^{\dagger},\delta B^{\dagger})^{T}$
and the input field vector $\mathcal{\mathscr{\mathcal{W}}}_{\text{in}}=(a_{\text{in}},c_{\text{in}}^{\dagger},B_{\text{in}}^{\dagger})^{T}$,
the linearized QLEs~(\ref{eq:BlueLinQLEs}) is written as
\begin{equation}
\frac{d\mathcal{\mathscr{\mathcal{W}}}}{dt}=-\mathscr{\mathcal{M}}\mathscr{\mathcal{W}}+\mathcal{J}\mathscr{\mathcal{W}}_{\text{in}}
\end{equation}
with the coefficient matrix 
\begin{equation}
\mathcal{\mathscr{\mathcal{M}}}=\left(\begin{array}{ccc}
i\Delta+\kappa_{a} & 0 & i\mathcal{G}_{a}\\
0 & -i\omega_{c}+\kappa_{c} & -iG_{c}\\
-i\mathcal{G}_{a}^{*} & -iG_{c} & -i\omega_{b}+\gamma_{B}
\end{array}\right).
\end{equation}
With the input-output relation, the output spectrum of the VIS mode
in the frequency domain is given by
\begin{equation}
S_{a,\text{out}}(\omega)=\mathcal{\mathscr{\mathcal{T}}}_{ac}^{\mathrm{bd}}(\omega)\left[S_{c,\text{in}}(\omega)+\frac{1}{2}\right]+S_{a,\text{add}}(\omega),
\end{equation}
where the conversion efficiency $\mathcal{\mathscr{\mathcal{T}}}_{ac}^{\mathrm{bd}}\left(\omega\right)$
from the IR signal to the VIS range is
\begin{align}
\mathcal{\mathscr{\mathcal{T}}}_{ac}^{\mathrm{bd}}\left(\omega\right) & =\left|\mathscr{\mathcal{U}}_{12}^{\mathrm{bd}}\left(\omega\right)\right|^{2}.\label{eq:BlueConEffic}
\end{align}
The added noise spectrum is $S_{a,\text{add}}(\omega)=\left[\mathcal{T}_{aa}^{\mathrm{bd}}(\omega)+\mathcal{T}_{aB}^{\mathrm{bd}}(\omega)\right]/2$,
where $\mathcal{\mathscr{\mathcal{T}}}_{aa}^{\mathrm{bd}}\left(\omega\right)=\left|\mathscr{\mathcal{U}}_{11}^{\mathrm{bd}}(\omega)\right|^{2}$
and $\mathcal{\mathscr{\mathcal{T}}}_{aB}^{\mathrm{bd}}\left(\omega\right)=\left|\mathscr{\mathcal{U}}_{13}^{\mathrm{bd}}(\omega)\right|^{2}$.
Here, $\mathscr{\mathcal{U}}_{1j}^{\mathrm{bd}}(\omega)$ is the element
in the first row and $j$-th column of the scattering matrix $\mathcal{\mathscr{\mathcal{U}}}^{\mathrm{bd}}(\omega)=\mathcal{\mathcal{J}}(\mathscr{\mathcal{M}}+i\omega I_{3})^{-1}\mathcal{\mathcal{J}}-I_{3}$,
and its specific component $\mathscr{\mathcal{U}}_{12}^{\mathrm{bd}}(\omega)$
is expressed as $\mathscr{\mathcal{U}}_{12}^{\mathrm{bd}}(\omega)=2G_{c}\mathcal{G}_{a}\sqrt{\kappa_{a}\kappa_{c}}/\{G_{c}^{2}(i\Delta+i\omega+\kappa_{a})+[(i\Delta+i\omega+\kappa_{a})(i\omega-i\omega_{b}+\gamma_{B})-\left|\mathcal{G}_{a}\right|^{2}](i\omega-i\omega_{c}+\kappa_{c})\}$.

\begin{figure}
\centering
\includegraphics[scale=0.8]{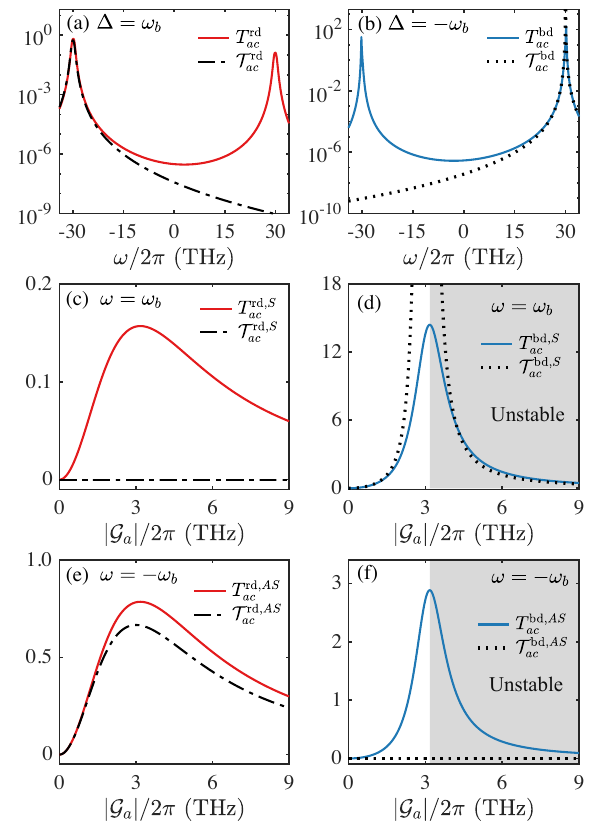}

\caption{(a, b) Conversion efficiencies $T_{ac}^{l}\left(\omega\right)$ and
$\mathcal{T}_{ac}^{l}\left(\omega\right)$ ($l=\text{rd},\text{bd}$,
denoting the cases of red-detuned and blue-detuned pump fields, respectively)
as functions of the frequency $\omega$ at $\left|\mathcal{G}_{a}\right|/2\pi=3\,\text{THz}$.
(c, d) Conversion efficiencies $T_{ac}^{l,S}$ and $\mathcal{T}_{ac}^{l,S}$
at the first Stokes sideband ($\omega=\omega_{b}$) as functions of
the coupling strength $\left|\mathcal{G}_{a}\right|$. (e, f) Conversion
efficiencies $T_{ac}^{l,AS}$ and $\mathcal{T}_{ac}^{l,AS}$ at the
first anti-Stokes sideband ($\omega=-\omega_{b}$) as functions of
the coupling strength $\left|\mathcal{G}_{a}\right|$. In panels (a,
c, e) {[}panels (b, d, f){]}, we use the red-detuned case of $\Delta=\omega_{b}$
(the blue-detuned case of $\Delta=-\omega_{b}$). Here we consider
the resonance case with $\omega_{c}=\omega_{b}=2\pi\times30\,\text{THz}$,
and other parameters are $N=10^{7}$, $g_{c}/2\pi=0.1\,\text{GHz}$,
$\varepsilon_{p}/2\pi=500\,\text{THz}$, $\kappa_{a}/2\pi=30\,\text{THz}$,
$\kappa_{c}/2\pi=0.5\,\text{THz}$, and $\gamma_{B}/2\pi=0.1\,\text{THz}$.
The gray shaded areas in panels (d) and (f) indicate the regions of
instability (without applying the RWA).}
\label{Fig4}
\end{figure}

In Fig.~\ref{Fig3}(b), we consider the blue-detuned pump field with
$\Delta=-\omega_{b}$ and the resonance case with $\omega_{c}=\omega_{b}$.
At the first anti-Stokes sideband of the pump field, i.e., $\omega_{p}+\omega_{b}$,
the conversion efficiency from the IR signal to the VIS range is given
by
\begin{equation}
\mathcal{\mathscr{\mathcal{T}}}_{ac}^{\mathrm{bd},AS}\equiv\mathcal{T}_{ac}^{\mathrm{bd}}(\omega=-\omega_{b})=\left|\frac{2G_{c}\mathcal{G}_{a}\sqrt{\kappa_{a}\kappa_{c}}}{G_{c}^{2}\Gamma_{a}-\left|\mathcal{G}_{a}\right|^{2}\Gamma_{c}+\Gamma_{a}\Gamma_{c}\Gamma_{B}}\right|^{2},\label{eq:BlueASConEff}
\end{equation}
where $\Gamma_{o}$ (for $o=a,c,B$) is defined below Eq.~(\ref{eq:RedSConEff}).
At the first Stokes sideband of the pump field, i.e., $\omega_{p}-\omega_{b}$,
the conversion efficiency of the IR-to-VIS signal is
\begin{equation}
\mathcal{\mathscr{\mathcal{T}}}_{ac}^{\mathrm{bd},S}\equiv\mathcal{T}_{ac}^{\mathrm{bd}}(\omega=\omega_{b})=\left|\frac{2G_{c}\mathcal{G}_{a}\sqrt{\kappa_{a}\kappa_{c}}}{G_{c}^{2}\kappa_{a}+\kappa_{a}\kappa_{c}\gamma_{B}-\left|\mathcal{G}_{a}\right|^{2}\kappa_{c}}\right|^{2}.\label{eq:BlueSConEff}
\end{equation}
By analyzing Eq.~(\ref{eq:BlueSConEff}), we find that the conversion
efficiency $\mathcal{\mathscr{\mathcal{T}}}_{ac}^{\mathrm{bd},S}$
from the IR signal to the VIS range reaches its maximum at the first
Stokes sideband when $\left|\mathcal{G}_{a}\right|^{2}\kappa_{c}\simeq G_{c}^{2}\kappa_{a}+\kappa_{a}\kappa_{c}\gamma_{B}$.

\section{Results}

Before discussing the impact of the added noise, we validate the RWA
and present in Fig.~\ref{Fig4} the conversion efficiency for IR-to-VIS
signal upconversion. Figures~\ref{Fig4}(a) and~\ref{Fig4}(b) show
the conversion efficiencies $T_{ac}^{l}(\omega)$ and $\mathcal{T}_{ac}^{l}(\omega)$
($l=\mathrm{rd},\mathrm{bd}$) as functions of the frequency $\omega$
for the red-detuned ($\Delta=\omega_{b}$) and blue-detuned ($\Delta=-\omega_{b}$)
pump fields, respectively. The solid line represents the numerical
result without using RWA $T_{ac}^{l}\left(\omega\right)$ in Eq.~(\ref{eq:ConEffic}).
The dash-dotted and dashed lines represent the analytical results
using RWA $\mathcal{T}_{ac}^{l}\left(\omega\right)$ calculated using
Eq.~(\ref{eq:RedConEffic}) and Eq.~(\ref{eq:BlueConEffic}), respectively.
The results show that, under the present parameters, the conversion
efficiencies $\mathcal{T}_{ac}^{l}$ and $T_{ac}^{l}$ are not in
complete agreement. In particular, the significant deviations occur
at the first Stokes (anti-Stokes) sideband of the red-detuned (blue-detuned)
pump field. However, within the frequency range of interest, i.e.,
the range of the first anti-Stokes (Stokes) of the red-detuned (blue-detuned)
pump field, the two conversion efficiencies agree qualitatively. For
the blue-detuned pump field, the conversion efficiency exceeds 1 at
both the first Stokes and anti-Stokes sidebands, indicating amplification
of IR-to-VIS signal upconversion.

Figures~\ref{Fig4}(c) and \ref{Fig4}(d) further show the conversion
efficiencies $T_{ac}^{l,S}$ and $\mathcal{T}_{ac}^{l,S}$ ($l=\text{rd},\text{bd}$)
at the first Stokes sideband as functions of the coupling strength
$\left|\mathcal{G}_{a}\right|$ for $\Delta=\omega_{b}$ and $\Delta=-\omega_{b}$,
respectively. It is shown that, $T_{ac}^{l,S}$ initially increases
with $\left|\mathcal{G}_{a}\right|$, reaches a maximum value (e.g.,
$T_{ac}^{\mathrm{rd},S,\text{max}}\approx0.13$ and $T_{ac}^{\mathrm{bd},S,\text{max}}\approx12$)
at optimal coupling strength $\left|\mathcal{G}_{a}\right|\simeq2\pi\times3.48\,\text{THz}$,
and then decreases. Similarly, Figures~\ref{Fig4}(e) and \ref{Fig4}(f)
display the conversion efficiencies $T_{ac}^{l,AS}$ and $\mathcal{T}_{ac}^{l,AS}$
at the first anti-Stokes sideband versus $\left|\mathcal{G}_{a}\right|$
for $\Delta=\omega_{b}$ and $\Delta=-\omega_{b}$, respectively.
Here, $T_{ac}^{l,AS}$ reaches its maximum value (e.g., $T_{ac}^{\text{rd},AS,\text{max}}\approx0.66$
and $T_{ac}^{\text{bd},AS,\text{max}}\approx2.39$) at optimal coupling
strength $\left|\mathcal{G}_{a}\right|\simeq2\pi\times3.48\,\text{THz}$.
Notably, under the red-detuned condition $\Delta=\omega_{b}$, the
anti-Stokes sideband exhibits a higher conversion efficiency than
the Stokes sideband, whereas the opposite behavior occurs under the
blue-detuned condition $\Delta=-\omega_{b}$. At the first anti-Stokes
(Stokes) sideband under the red- (blue-) detuned condition, the conversion
efficiencies $\mathcal{T}_{ac}$ and $T_{ac}$ begin to deviate for
the coupling strength $\left|\mathcal{G}_{a}\right|\gtrsim2\pi\times1\,\text{THz}$,
as shown in Figs.~\ref{Fig4}(d) and \ref{Fig4}(e). Nevertheless,
the qualitative physical picture remains consistent, especially regarding
the resonance peak positions. Based on these findings, we next analyze
the conversion efficiency and the added noise without applying the
RWA.

\begin{figure}
\centering
\includegraphics[scale=0.8]{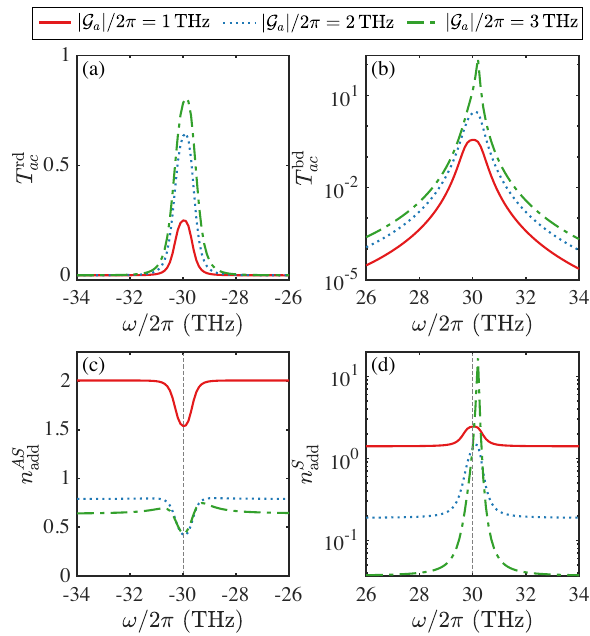}

\caption{(a, c) Conversion efficiency $T_{ac}^{\text{rd}}(\omega)$ and added
noise $n_{\text{add}}^{AS}(\omega)$ at the first anti-Stokes sideband,
and (b, d) Conversion efficiency $T_{ac}^{\text{bd}}(\omega)$ and
added noise $n_{\text{add}}^{S}(\omega)$ at the first Stokes sideband,
as functions of the frequency $\omega$. The results correspond to
the red-detuned ($\Delta=\omega_{b}$) and blue-detuned ($\Delta=-\omega_{b}$)
cases, respectively, with coupling strength $\left|\mathcal{G}_{a}\right|/2\pi=1,2,3\,\text{THz}$.
Here we consider the resonance case of $\omega_{c}=\omega_{b}=2\pi\times30\,\text{THz}$,
and other parameters are the same as those in Fig.~\ref{Fig4}.}
\label{Fig5}
\end{figure}

We now discuss the impact of the added noise. Figures~\ref{Fig5}(a)
and \ref{Fig5}(c) show the conversion efficiency $T_{ac}^{\text{rd}}(\omega)$
and the added noise $n_{\text{add}}^{AS}(\omega)$ at the first anti-Stokes
sideband as functions of the frequency $\omega$, under the red-detuned
condition $\Delta=\omega_{b}$ for different coupling strengths $\left|\mathcal{G}_{a}\right|$.
As shown in Figs.~\ref{Fig5}(a, c), the peak value of the conversion
efficiency increases with $\left|\mathcal{G}_{a}\right|$, while the
added noise decreases. Notably, at large $\left|\mathcal{G}_{a}\right|$,
the added noise on resonance ($\omega=-\omega_{b}$) approaches half
a quantum, i.e., $n_{\text{add}}^{AS}\rightarrow1/2$. In contrast,
Figures~\ref{Fig5}(b) and \ref{Fig5}(d) present the conversion
efficiency $T_{ac}^{\text{bd}}(\omega)$ and the added noise $n_{\text{add}}^{S}(\omega)$
at the first Stokes sideband under the blue-detuned condition $\Delta=-\omega_{b}$
for $\left|\mathcal{G}_{a}\right|/2\pi=1,2,3\,\text{THz}$. With increasing
coupling strength $\left|\mathcal{G}_{a}\right|$, the conversion
efficiency increases with $\left|\mathcal{G}_{a}\right|$, whereas
the added noise exhibits a non-monotonic behavior. At $\left|\mathcal{G}_{a}\right|/2\pi=2\,\text{THz}$,
the added noise on resonance ($\omega=\omega_{b})$ reaches 1.3.

\begin{figure}
\centering
\includegraphics[scale=0.8]{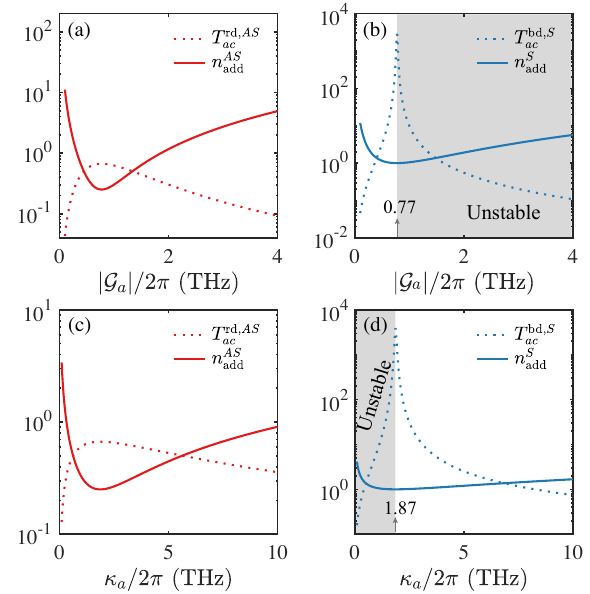}

\caption{Conversion efficiency $T_{ac}^{\text{rd},AS}$ and added noise $n_{\text{add}}^{AS}$
on resonance at the first anti-Stokes sideband as functions of (a)
$\left|\mathcal{G}_{a}\right|$ for $\kappa_{a}/2\pi=2\,\text{THz}$
and (c) $\kappa_{a}$ for $\left|\mathcal{G}_{a}\right|/2\pi=0.75\,\text{THz}$
under the red-detuned condition $\Delta=\omega_{b}$. Conversion efficiency
$T_{ac}^{\text{bd},S}$ and added noise $n_{\text{add}}^{S}$ on resonance
at the first Stokes sideband as functions of (b) $\left|\mathcal{G}_{a}\right|$
for $\kappa_{a}/2\pi=2\,\text{THz}$ and (d) $\kappa_{a}$ for $\left|\mathcal{G}_{a}\right|/2\pi=0.75\,\text{THz}$
under the blue-detuned condition $\Delta=-\omega_{b}$. Other parameters
are the same as those in Fig.~\ref{Fig4}.}
\label{Fig6}
\end{figure}

We further analyze how the system parameters--$\left|\mathcal{G}_{a}\right|$
and $\kappa_{a}$--affect both the conversion efficiency and added
noise. Figures~\ref{Fig6}(a, c) show the conversion efficiency $T_{ac}^{\text{rd},AS}$
(dotted curves) and the added noise $n_{\text{add}}^{AS}$ on resonance
(solid curves) at the first anti-Stokes sideband under the red-detuned
condition $\Delta=\omega_{b}$. As illustrated in Fig.~\ref{Fig6}(a,
c), the conversion efficiency first increases and then decreases with
$\left|\mathcal{G}_{a}\right|$ and $\kappa_{a}$, while the added
noise shows the reverse trend. By tuning the coupling strength and
decay rate of the system, the added noise can be suppressed to a level
below half a quantum; however, the conversion efficiency $T_{ac}^{\text{rd},AS}\lesssim1$.
In Figs.~\ref{Fig6}(b, d), we show the conversion efficiency $T_{ac}^{\text{bd},S}$
(dotted curves) and the added noise $n_{\text{add}}^{S}$ on resonance
(solid curves) at the first Stokes sideband under the blue-detuned
condition $\Delta=-\omega_{b}$. In the stable regime, the conversion
efficiency increases with $\left|\mathcal{G}_{a}\right|$ but decreases
with $\kappa_{a}$, while the added noise shows the reverse trend.
In addition, at $\left|\mathcal{G}_{a}\right|=2\pi\times0.75\,\text{THz}$
and $\kappa_{a}/2\pi=2\,\text{THz}$, we find that the conversion
efficiency reaches $T_{ac}^{\text{bd},S}\approx10^{3}$, while the
added noise approaches the quantum limit of one quantum, i.e., $n_{\text{add}}^{S}\rightarrow1$.
Compared with the red-detuned case, the added noise increases by 2
to 3 times, while the conversion efficiency is amplified by three
orders of magnitude.

\section{Conclusion}

In conclusion, we have calculated the conversion efficiency from IR
signal to VIS range and the noise added to the upconverted signal
in a molecular optomechanical cavity, where the molecular vibrations
are coupled to both the IR and VIS modes. Our results indicate that,
under the red-detuned condition, the conversion efficiency of the
anti-Stokes sideband is higher than that of the Stokes sideband, whereas
the conversion efficiency of the Stokes sideband surpasses that of
the anti-Stokes sideband under the blue-detuned condition, where amplification
of the IR signal is achieved. In addition, we find that the added
noise remains below half a quantum under the red-detuned condition,
while it approaches the quantum limit of one quantum under the blue-detuned
condition. Compared with the red-detuned regime, the added noise in
the blue-detuned regime increases by a factor of 2 to 3, while the
conversion efficiency is amplified by three orders of magnitude. Our
research seeks to establish a theoretical basis to support experimental
progress in IR signal upconversion.

\begin{acknowledgments}
This work is supported by the Quantum Science and Technology-National
Science and Technology Major Project (Grant No.~2023ZD0300700), the
National Natural Science Foundation of China (Grants No.~12574387,
No.~12405011, and No.~U2230402), and the Natural Science Foundation
of Hainan Province (Grants No. 125QN210 and No. 125RC631).
\end{acknowledgments}


\begin{thebibliography}{66}
\bibitem [{not()}]{note}%
	\BibitemOpen
	\href@noop {} {\bibinfo {title} {{T}he correspondence between the first
			anti-stokes (stokes) sideband of the pump field and $\omega = -\omega_{b}$
			($\omega = \omega_{b}$) arises from two considerations. (i) {T}o eliminate
			the time-dependent terms in {H}amiltonian (1), we transform the system into
			the interaction picture with respect to $\omega_{p}a^{\dagger}a$. (ii) {T}he
			{F}ourier transform is defined as in {E}q. (10), where the exponential factor
			takes the form $e^{-i\omega t}$.}}\BibitemShut {Stop}%
\end{thebibliography}
%

\end{document}